\documentstyle[preprint,aps,prl]{revtex}
\tightenlines
\begin{document}
\draft
\bibliographystyle{prsty}
\title{ANGULAR FORCES AROUND TRANSITION METALS\\
IN BIOMOLECULES}
\author{A. E. Carlsson}
\address{Department of Physics, CB 1105\\
Washington University, St. Louis, MO 63130}
\date{Submitted to Physical Review Letters, \today}
\maketitle
\begin{abstract}
Quantum-mechanical analysis based on an exact sum rule
is used to extract an semiclassical angle-dependent energy 
function for transition metal ions in biomolecules. The angular
dependence is simple but different from existing classical
potentials. Comparison of predicted energies with 
a computer-generated database shows that the semiclassical 
energy function is remarkably accurate, and that its 
angular dependence is optimal.
\end{abstract}

Biomolecular modeling with classical potentials has become an 
increasingly important tool in problems such as
the determination of protein structure and function, 
and the design of new molecules with desired properties.
With the continuing availability of increasingly powerful computers, 
one can only expect this growth to continue. 
A major roadblock toward expanded use of modeling with classical 
potentials is the absence of sufficiently reliable force laws 
for transition metals in biomolecules. 
The ability to treat transition metals is important because the active
sites of many proteins are defined by transition metals; also smaller
biologically active molecules often have transition metals as crucial
constituents. In fact, transition metal complexes have been
proposed as crucial ingredients in the origin of 
life itself\cite{Wachtershauser97}.
Unlike the $s$-$p$ constituents of biomolecules, which usually
have unique, well-defined bonding configurations
(such as $sp^2$ planar coordination), transition metals can adopt a
broad range of asymmetric environments. 
This asymmetry is often important for the functioning of enzymes.  
Thus, for biomolecular modeling, one needs a ``generic'' potential 
which treats essentially all physically reasonable environments instead 
of perturbations relative to as single well-defined structure. 
{\it A priori}, one does not know the functional form of such a generic 
potential. The pair approximation, which ignores angular constraints,
is applicable to simple metal ions, but not to transition metals. 
The transition-metal $d$-orbitals lead to complex angular forces 
which are manifested, for example, in the frequent occurrence of 
Cu$^{2+}$ and Ni$^{2+}$ ions in square-planar environments that 
are unexpected on the basis of pair interactions alone. 
In existing simulation codes based on classical potentials, 
the angular terms are usually either 
ignored\cite{Hancock90,Bernhardt92,Cundari96}, on the assumption 
that direct ligand-ligand interactions can take up most of the 
``slack'', or they are treated with simple assumed angular forms.  
The latter range from quadratic or higher order expansions 
about observed equilibrium bond 
angles\cite{Wiesemann94,Timofeeva95,Halgren96,Geremia97}
to more sophisticated expansions in trigonometric 
functions\cite{Mayo90,Allured91,Rappe93,Comba95}.
However, there has been no derivation of the angular form of 
classical potentials from quantum mechanics. 

In this Letter, I use quantum-mechanical analysis to derive an energy 
function for $d$-electrons based on the local environment in 
biomolecules.  The energy function has a ``semiclassical'' form, 
in the sense that it is slightly more complex than a classical additive 
sum of ligand-ligand interactions, but is still straightforward to treat 
in molecular modeling codes. To test the energy function, 
I generate a large number of random transition metal 
environments and evaluate their exact energies as a test set. 
The $d$-electron energy is described with surprising precision. 
The accuracy is much better than that of commonly used functional forms, 
and significantly improves on that of additive energy functions. 
The angular dependence of an energy function obtained by fitting 
to the exact energies is very similar to that derived analytically.

In biomolecules, transition metals are typically in a ``coordination''
bonding configuration. This differs from metallic bond in 
elemental transition metals in that the $d$-states usually hybridize with
ligand orbitals at lower energies, rather than other 
$d$-orbitals at the same energy. This leads to well-defined discrete
charge states. The physics of coordination bonding is well
described by the ligand field theory\cite{Gerloch81} (LFT), 
which treats the $d$-shell in a transition-metal ligand complex 
by an effective $d$-$d$ Hamiltonian:
\begin{equation}
 {\widehat H}_d = \sum_{\mu ,\nu } h_{\mu\nu}
                  |d_{\mu}\rangle \langle d_{\nu}|\quad .
\label{hd}
\end{equation}
Here $|d_{\mu}\rangle$ and $|d_{\nu}\rangle$ are $d$-basis orbitals
on the transition-metal ion, and the $h_{\mu \nu }$ 
contain the effects of the ligands in a perturbative fashion:
\begin{equation}
  h_{\mu \nu}
= \sum_i \langle d_{\mu}|H|i\rangle
  \langle i|H|d_{\nu}\rangle /[E_d -E_i]\quad ,
\label{hmunu}
\end{equation}
where the $|i\rangle$ are orbitals on the ligands that hybridize with
the $d$-shell, and $E_d$ and $E_i$ are the $d$-shell and ligand-orbital
energies, respectively. (The ligands are all taken to be equivalent for
simplicity, but the more general cases are treated straightforwardly.)
This approximate treatment describes
the systematics of $d$-shell splittings in transition metal complexes
quite well, although the electronic transition energies are not obtained
quantitatively. In the case where only $\sigma$-type interactions between
the ligands and the $d$-shell are present, the matrix elements of the
effective Hamiltonian can be written\cite{Gerloch81} as
\begin{equation}
  h_{\mu \nu}
= \sum_i e (r_i) Y_{\mu} ({\hat r}_i)Y_{\nu}({\hat r}_i ).
\label{aom}
\end{equation}
where the $Y_{\nu}$ have the angular dependence of the $d$-basis
orbitals, and the radial function $e(r_i )$ includes the 
effects of the energy denominator as well as the matrix elements.

The $d$-electron energy associated with ${\widehat H}_d$ is obtained 
by simply adding the eigenenergies of the occupied $d$-states. 
This approximation is justified when comparing structural energies 
within a single well-defined charge/spin state.  We focus on the 
``ligand-field stabilization energy'' 
$E_{\rm LFSE}=\sum_n\varepsilon_n-N_d\bar\varepsilon$. 
Here the first term denotes the eigenvalue sum, $N_d$ is the number 
of $d$-electrons, and $\bar\varepsilon$ is the average energy of the 
$d$-complex (including both occupied and unoccupied states). 
As indicated in Fig.~1, splitting of the $d$-complex by ligand-field
interactions provides a negative (stabilizing) contribution
to $E_{\rm LFSE}$ if the $d$-shell is partly filled. The stabilizing
contribution is enhanced if there is a gap between the highest 
occupied and lowest unoccupied states, as occurs for Cu$^{2+}$ 
and Ni$^{2+}$ ions in the square-planar coordination.  
We define the half-width $W$ of the $d$-complex as the rms deviation 
of the energy eigenvalues from the $d$-complex average energy 
$\bar\varepsilon$. In the first approximation, one expects that
$E_{\rm LFSE}$ should be proportional to $W$.

The $d$-electron energy function developed here gives $E_{\rm LFSE}$ as
a simple function of the ligand positions. It is
based on an exact sum rule that for $W$. Explicit calculation
via Eqs.~(\ref{hd}) and (\ref{hmunu}) shows that
\begin{equation}
5W^2 = \sum_n (\varepsilon_n -\bar\varepsilon )^2
     = {\rm Tr} ({\widehat H}_d - \bar\varepsilon \hat I )^2 
     = \sum_{i,j} U_{ij}\, , 
\end{equation}
where the ligand-ligand interaction is defined by
\begin{eqnarray}
U_{ij} = && \left( \sum_\mu \langle i|H|d_{\mu}\rangle 
            \langle d_{\mu}|H|j\rangle \right)^2/(E_d-E_i)(E_d-E_j)\nonumber\\ 
       - &&~(1/5)\left[ \sum_{\mu} \langle i|H|d_{\mu}\rangle
            \langle d_{\mu}|H|i\rangle /(E_d-E_i)\right] \nonumber \\
  \times &&~\left[ \sum_{\nu} \langle j|H|d_{\nu}\rangle
            \langle d_{\nu}|H|j\rangle /(E_d-E_j)\right]
\label{Uij}
\end{eqnarray}

For the case described by Eq.~(\ref{aom}), the interaction takes the form 
\begin{equation}
   U_{ij} = e(r_i) e(r_j) 
            \Bigl[ P_2 (\cos{\theta_{ij}})^2-(1/5)\Bigr] 
\label{Uijsigma}
\end{equation}
where $P_2(\theta )=(3\cos^2\theta -1)/2$ is the second-order
Legendre polynomial. This, and the assumption that 
$E_{\rm LFSE}$ is proportional to $W$, motivates the following choice 
for the functional form of $E_{\rm LFSE}$ in terms of the local environment:
\begin{equation}
    E_{\rm LFSE} 
  = -\left[ \sum_{ij} e(r_i) e(r_j) u(\theta_{ij})\right]^{1/2}
\label{elfse}
\end{equation}
where 
\begin{equation}
   u(\theta ) = \bigl[ P_2 (\cos{\theta_{ij}})^2-(1/5)\bigr]\, .
\label{u}
\end{equation}
Because the square root of the ligand-ligand sum is taken, 
this type of energy function is different from classical
additive angular interaction potentials. 
I call it a ``semiclassical'' energy function, 
since the steps in its calculation are similar to those in 
the calculation of a classical energy function, 
but quantum mechanical effects are included in a systematic fashion.
It applies to one spin component of a transition metal $d$-shell;
if both spin components contribute, then $E_{\rm LFSE}$
is simply the sum of contributions from the two components. 
The form (\ref{elfse}) is parallel in form to 
``many-atom''\cite{Finnis84} and ``embedded-atom''\cite{Daw83} 
energy functions, but these are not angle-dependent. 
Modifications of the embedded atom method\cite{Baskes92} have
included angular dependence, but without quantum-mechanical
grounding, assuming angular forms very different from the
present ones. 

In order to evaluate the accuracy of this functional form in the types 
of disordered local geometries that may be found in biomolecular
environments, I have evaluated exact cluster energies (from the
eigenvalues of Eq.~(\ref{hd})) for an ensemble of transition-metal
complexes having random bond lengths and angles. 
The transition metal ions have six neighboring ligands.  
The coupling strengths $e_i=e(r_i)$ in Eqs.~(\ref{aom}) 
and (\ref{elfse}) vary randomly between 0 and 2 (in arbitrary units), 
corresponding to distances varying from a short-range cutoff to infinity, 
and the orientations $\hat r_i$ are chosen at random.
In this way, a very broad range of environments, with effectively varying 
coordination numbers, is sampled. Semiclassical energy functions of 
the form (\ref{elfse}), as well as classical energy functions, have been
least-squares fitted to the exact $d$-electron energies of these clusters,
for the ions Fe$^{2+}$ through Cu$^{2+}$, taken in the high-spin 
configuration (Mn$^{2+}$ and Zn$^{2+}$ are not included, since 
their minority-spin $d$-bands are empty and filled respectively, 
so $E_{\rm LFSE}$ vanishes).
In the fits, in addition to the ligand-ligand interaction terms, 
we include a constant term in the ligand-ligand interaction,
as well as a sum of single-ligand terms.  Figure~2a shows the fit for 
Cu$^{2+}$ obtained with the semiclassical energy function (\ref{elfse}). 
The energies are fit remarkably well, with the standard deviation
of 0.16 being less than 10 percent of the typical values of 
$|E_{\rm LFSE}|$. Similar results are obtained for Ni. 
For Co$^{2+}$, the fractional error is about 15 percent. 
For Fe$^{2+}$, the magnitude of $E_{\rm LFSE}$ is found to be 
an order of magnitude smaller than for Cu$^{2+}$ and Ni$^{2+}$, 
and the fractional error resulting from using the potentials 
is about 50 percent; nevertheless, the absolute errors are about 
half of those for Cu$^{2+}$ and Ni$^{2+}$.
Figure~2b shows corresponding results for a classical potential for
Cu$^{2+}$ of the form $\sqrt{e_i}\sqrt{e_j}\sin^2{2\theta}$, 
where the angular dependence is taken from recent simulations 
of cluster energetics\cite{Comba95} and the $\sqrt{e_i}$ dependence
follows from dimensional analysis and the linear scaling
of $E_{\rm LFSE}$ with uniform scaling the $e_i$. We take this
form to be typical of the treatment of transition metals in standard 
modeling packages in which simple plausible forms are assumed.  
The fit is much less accurate, with a standard deviation of 0.49. 

The energy function (\ref{elfse}-\ref{u}) provides 
an optimal description of the $d$-shell energetics 
in terms of two-ligand interaction interactions. To
show this, I have fitted more elaborate
potentials of the form (\ref{elfse}) to the energy database,
in which $u(\theta )$ is represented by a sum of terms of
the form $\cos{n \theta}$, with $n \le 8$. The results
are shown in Fig.~3. The agreement between the optimized
$u(\theta )$ and the form (\ref{u}) is almost exact for
Cu$^{2+}$ and Ni$^{2+}$, and very good for Co$^{2+}$. For
Fe$^{2+}$, the absolute discrepancies are small, but the
relative discrepancies are larger.  Note that the shapes of 
$u(\theta )$ as obtained here differ completely from the 
$\sin^2{2\theta }$ form of Ref.~\onlinecite{Comba95}, which
is shown by the dotted curve in frame (a).  
In addition, I have tried modified forms of Eq.~(\ref{elfse}),
in which the square root is replaced by a power law dependence,
so that an exponent of unity gives an additive potential. The
minimum error is obtained with an exponent very close to 0.50,
corresponding to the Eq.~(\ref{elfse}).  Thus we have 
fairly definitively pinned down the functional form of the
angular forces around these ions. We note that these
results are also applicable to the low-spin versions of
the ions, by simple addition of contributions from the two
subbands. Then, for example, low-spin Ni$^{2+}$ becomes 
equivalent to high-spin Cu$^{2+}$.

The main chemical trend in $u(\theta )$ with changing $d$-count 
is a change in the magnitude of the potential, rather than its shape. 
The potentials for Ni$^{2+}$ and Cu$^{2+}$ are similar in magnitude, 
that for Co$^{2+}$ roughly a factor of two weaker, and that for 
Fe$^{2+}$ is weaker by an order of magnitude. The weakness of the 
Fe$^{2+}$ potential can be partly understood by analysis of the 
energetics of four-ligand complexes. For these, ${\widehat H}_d$, 
as given in Eq.~(\ref{hd}), is a sum of four one-dimensional projection 
operators thus has rank four. One readily shows that all of its eigenvalues 
are nonnegative.  This means that the lowest eigenvalue is zero, 
independent of the angular arrangement of the ligands.  In the case 
of Fe$^{2+}$, there is only one $d$-electron, which resides in the 
orbital having the zero eigenvalue. Thus there are no angular interactions 
for Fe$^{2+}$ with four ligands. For cases with higher coordination, 
the lowest eigenvalue will still likely be close to zero 
unless the five contributing projection operators are orthogonal 
to each other.  From the point of view of practical application, 
the variations seen in Fig.~3 suggest that the inclusion of angular 
forces for modeling Cu$^{2+}$ and Ni$^{2+}$ is crucial, but that 
the Fe$^{2+}$ ion (in high-spin configuration) might well be modeled 
with only radial interactions.

These features can be used to explain the observed chemical trends in the 
relative stability of square and tetrahedral structures in these systems.
I have evaluated the energy difference $\Delta E$ between $E_{\rm LFSE}$ 
between the square and tetrahedral coordination geometries. Comparisons 
between the exact values and those obtained by Eq.~(\ref{elfse}) 
and the empirical potential\cite{Comba95} are shown in Fig.~4, 
for the transition metal ions Fe$^{2+}$ through Cu$^{2+}$. 
The empirical-potential results are much too small, but the basic 
trends of the exact results are also seen in the semiclassical results, 
with the square structure favored strongly for Ni$^{2+}$ and Cu$^{2+}$.
This trend is consistent with known structures of four-ligand 
transition metal complexes. Such complexes of Ni$^{2+}$ and Cu$^{2+}$ 
overwhelmingly adopt square coordination, in the absence of
steric constraints, while Fe$^{2+}$ and Co$^{2+}$ 
generally have tetrahedral coordination\cite{Zelewsky96,Foot2}.
(We note that the experiments do not necessarily establish 
the sign of the electronic contribution $\Delta E$ calculated 
here for a given system, since direct electrostatic interactions 
between the ligands tend to favor tetrahedral coordination; 
only the trend with varying $d$-count is established.)
The structural energies can be understood with the help of the potentials 
shown in Fig.~3. The minima at $0^\circ$ and $180^\circ$ favor the square 
structure in all cases, but are weaker for Fe$^{2+}$ and Co$^{2+}$. 
In fact, the calculated values of $\Delta E$ correspond fairly closely 
to the strengths of the angular interactions.  The energy differences 
are not, however, obtained quantitatively by the semiclassical energy 
function.  The discrepancy lies mainly in the energy of the tetrahedral 
structure. For tetrahedral  Co$^{2+}$, for example, the semiclassical 
energy function underestimates $|E_{\rm LFSE}|$ by about 20 percent. 

In summary, I have shown that a new semiclassical angular energy function,
with a simple analytic angular dependence, describes the ligand-field 
stabilization energy for transition-metal ions in biomolecules remarkably 
well.  The theoretical form for the angular dependence is strongly confirmed 
by analysis of a large computer-generated database of complexes.  
Analysis of the angular form of the interactions justifies the systematics 
of the relative stability of square and tetrahedral packing in terms 
of the behavior of the interactions at $0^\circ$ and $180^\circ$. 
Incorporation of this form of energy function into existing biomolecular 
simulation packages should significantly enhance their reliability, 
and lead to new possibilities for design of metal-containing biomolecules.

\acknowledgments

This work received support from the Department of Energy under
Grant number DE-FG02-84ER45130, which is gratefully acknowledged.


\begin{thebibliography}{10}
\bibitem{Foot2}
Some of the structures are for low-spin rather than high-spin configurations.
  For Ni$^{2+}$, as discussed in the text, the the low-spin configuration leads
  to an energy function with the same form as for Cu$^{+2}$. Low-spin Fe$^{2+}$
  and Co$^{2+}$ generally have tetrahedral cooordination.
\end{thebibliography}

\begin{figure}
\caption{Ligand-field splitting of Ni$^{2+}$ $d$-shell in square 
coordination. Only minority spin band, in high-spin
configuration, is shown.}
\end{figure}

\begin{figure}
\caption{Accuracy test of semiclassical and empirical energy
functions, in comparison with exact quantum-mechanical results
for ligand-field Hamiltonian. Energy unit is average coupling
of single ligand to transition-metal $d$-shell.}
\end{figure}

\begin{figure}
\caption{Angular dependence of energy function. Solid lines:
ten-parameter fit to exact energies. Dashed lines: derived
angular function from Eq. (8).  Function $u(\theta )$ is dimensionless.  
Frame (a) Cu$^{2+}$; (b) Ni$^{2+}$; (c) Co$^{2+}$; (d) Fe$^{2+}$. 
Dotted line in frame (a) is empirical energy function from Ref. [12], 
with magnitude adjusted for clear comparison.}
\end{figure}

\begin{figure}
\caption{Energy differences $\Delta E$ between square and tetrahedrally
coordinated transition metal ions. Energy unit is coupling strength 
between single ligand and transition metal. Solid circles: exact treatment 
of ligand-field Hamiltonian.  Open circles: semiclassical energy function 
(Eq. (7)).  Triangles: empirical energy function (Ref. [12]).}
\end{figure}

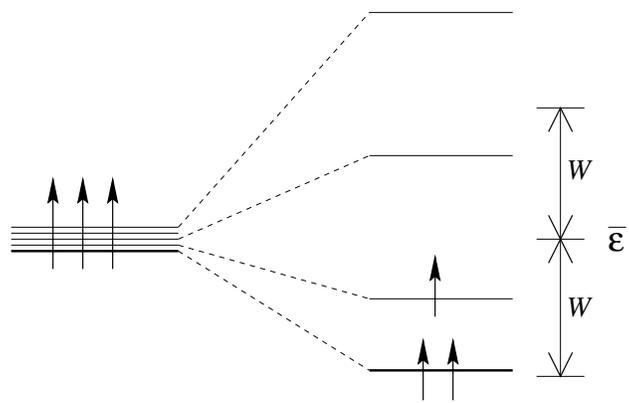
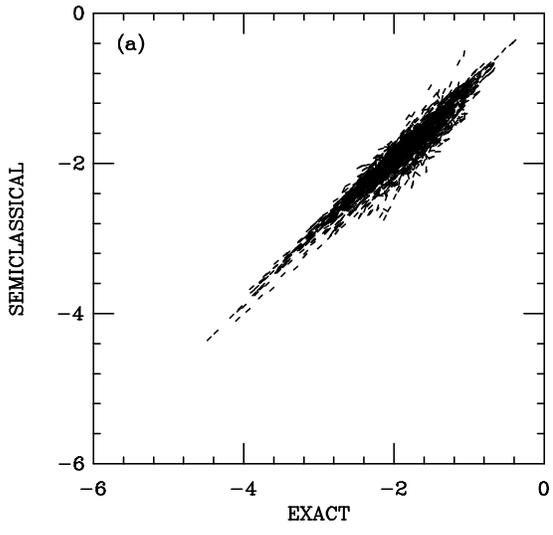
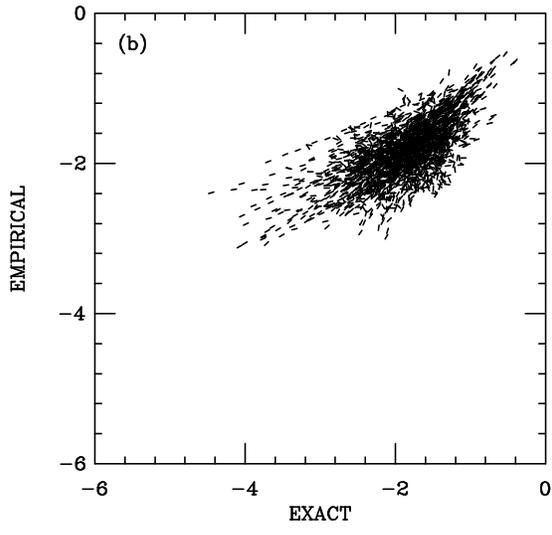
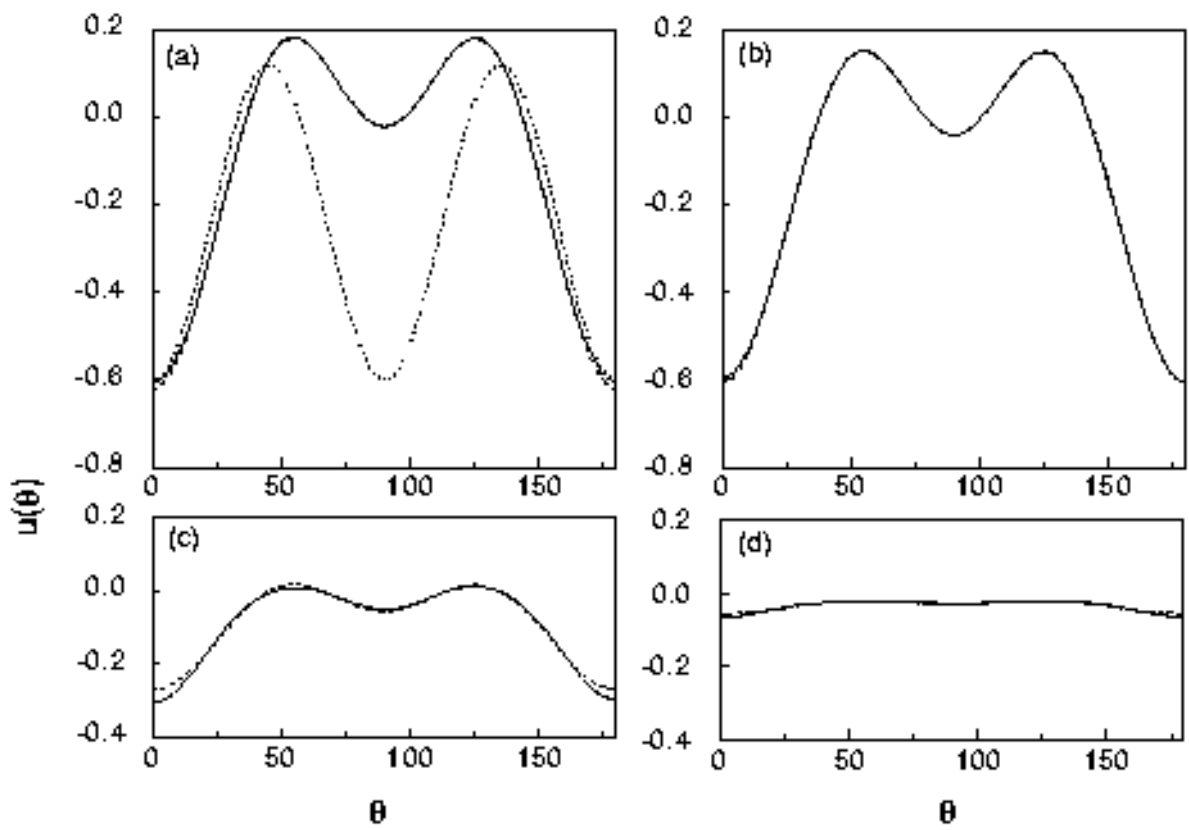
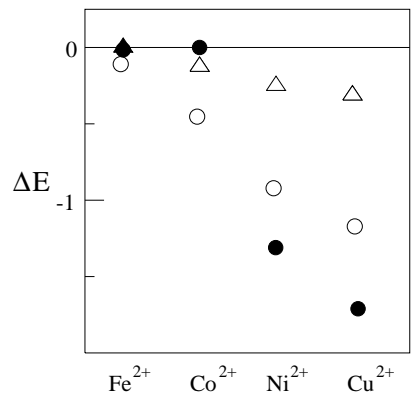
\end{document}